\documentclass[12pt] {article}
\usepackage{amsmath, epsfig}
\title{A Quantum Field Theory Term Structure Model Applied to Hedging}
\author{Belal E. Baaquie, Marakani Srikant, and Mitch Warachka}
\date{}

\newtheorem{thm}{Theorem}[section]
\newtheorem{rem}{Remark}[section]
\newtheorem{define}{Definition}[section]
\newtheorem{lem}{Lemma}[section]
\newtheorem{cor}{Corollary}[section]
\newtheorem{prop}{Proposition}[section]
\begin{document}
\maketitle

\begin{abstract}
  A quantum field theory generalization, Baaquie \cite{Baaquie}, of the Heath, Jarrow, and Morton (HJM) \cite{HJM} term structure model parsimoniously describes the evolution of imperfectly correlated forward rates. Field theory also offers powerful computational tools to compute path integrals which naturally arise from all forward rate models. Specifically, incorporating field theory into the term structure facilitates hedge parameters that reduce to their finite factor HJM counterparts under special
correlation structures. Although investors are unable to perfectly hedge against an infinite number of term structure perturbations in a field theory model, empirical evidence using market data reveals the effectiveness of a low dimensional hedge portfolio.
\end{abstract}

\section{Introduction}

Applications of physics to finance are well known, and the application of quantum mechanics to the theory of option pricing is well known. Hence it is natural to utilize the formalism of  quantum field theory to study the evolution of forward rates. Quantum field theory models of the term structure originated with Baaquie \cite{Baaquie} and the hedging properties of such models are analyzed in this paper. The intuition behind quantum field theory models of the term structure stems from allowing each forward rate maturity to both evolve randomly and be imperfectly correlated with every other maturity. As pointed out in Cohen and Jarrow \cite{Cohen}, this may also be accomplished by increasing the number of random factors in the original HJM towards infinity. However, the infinite number of factors in a field theory model are linked via a single function that governs the
correlation between forward rate maturities. Thus, instead of estimating additional volatility functions in a multifactor HJM framework, one additional parameter is sufficient for a field theory model to instill imperfect correlation between every forward rate maturity. As the correlation between forward rate maturities approaches unity, field theory models reduce to the standard one\footnote{A field theory model may also converge to a multifactor HJM model as illustrated in Proposition \ref{Con-HJM}.} factor HJM model. Therefore, the fundamental difference between finite factor HJM and field theory models is the minimal structure the latter requires to instill imperfect correlation between forward rates. Section \ref{Est} demonstrates the challenge to reliably estimating volatility functions in a multifactor HJM model. However, it should be stressed that field theory models originate from the finite factor HJM framework with the Brownian motion(s) replaced by a field. As seen in the next section, crucial results such as the forward rate drift restriction are generalized but remain valid for an infinite factor process.

The imperfect correlation between forward rates in a field theory
model addresses the theoretical and practical challenges posed by
their finite factor counterparts. Despite the enormous contribution of
the finite factor HJM methodology, an important dilemma is
intrinsic to the finite factor framework. For a finite factor model to
correctly match the movements of $N$ forward rates, an $N$ factor
model is required whose volatility parameters are difficult to
estimate once $N$ increases beyond one, Amin and Morton
\cite{Morton} and section \ref{Est}. However, field theory
models only require an additional parameter to describe the
correlation between forward rate maturities. Consequently, field
theory models offer a parsimonious methodology to ensure the evolution
of forward rates is consistent with the initial term structure without
relying on a improbably large number of factors or eliminating the
possibility of certain term structures, Bj\"{o}rk and Christensen
\cite{Con}.  More precisely, a field theory model does not permit a finite linear combination of forward rates to exactly describe the innovation in
every other forward rate. From a practical perspective, an $N$ factor term structure model implies the existence of an $N$ dimensional basis for forward rates that allows $N$ bonds with distinct maturities to be arbitrarily chosen when hedging. However, in the context of a field theory model, hedging performance depends on the maturity of the bonds in the hedge
portfolio as illustrated in subsection \ref{error}. To summarize, the
primary advantage of field theory models is their ability to generate
a wider class of term structure innovations in a parsimonious manner.
Consequently, hedging performance depends on the constituents of the
hedge portfolio. This important issue is addressed empirically in section \ref{Est} after the derivation of hedge parameters in section \ref{h}. As a final observation, stochastic volatility in a finite factor HJM model is implicitly a special case of field theory.

Assigning independent random variables to each forward rate maturity
has been studied by utilizing stochastic partial differential
equations involving infinitely many variables. Past research that instilled imperfect correlation between forward rates with generalized continuous random processes includes Kennedy \cite{Ken}, Kennedy \cite{Ken1}, Goldstein \cite{Gold}, Santa Clara and Sornette \cite{String}. The field theory approach developed by Baaquie \cite{Baaquie} is complimentary to past research since the expressions for all financial instruments are formally given as a functional integral. However, a significant advantage offered by field theory is the variety of
computational algorithms available for applications involving pricing
and hedging fixed income instruments. In fact, field theory was
developed precisely to study problems involving infinitely many
variables and the path integral described in the next section which serves as a generating function for forward curves is a natural tool to employ in  applications of term structure models. To date, there has been little empirical testing of previous field theory models which is understandable given their incomplete structure. The purpose of this paper is to demonstrate the practical advantages of utilizing field theory when hedging; specifically, the computational power offered by the path integral not found in previous research.

Although field theory term structure models offer several improvements
over finite factor models, past research has not fully exploited all
simplifying aspects of utilizing field theory. Specifically,
implementation of field theory models when hedging bonds requires
generalizing the concept of hedging. Since a field theory model may be
viewed as an infinite factor model, perfectly hedging a bond portfolio
is impossible\footnote{A measure valued trading strategy as in
Bj\"{o}rk, Kabanov, and Runggaldier \cite{BKR} provides one
alternative.} although computing hedge parameters
remains feasible. Furthermore, empirical evidence in section \ref{Est}
finds the correlation between forward rate maturities exerts a
significant impact on the hedging of bonds and allows a low
dimensional basis to effectively hedge term structure risk. It is
important to emphasize that field theory facilitates rather than
inhibits the implementation of term structure models as seen when the
path integral for forward rates is introduced. The path integral is a
powerful theoretical tool for computing expectations involving forward rates
required for pricing contingent claims such as futures contracts in a straightforward manner. As expected, closed form solutions for fixed income contingent claims and hedge parameters reduce to standard ``textbook" solutions such as those found in Jarrow and Turnbull \cite{JTtext} when the correlation between forward rates approaches unity. In summary, this paper
elaborates on an implementable field theory term structure model that
addresses the limitations inherent in finite factor term structure
models.

This paper is organized as follows. Section two briefly summarizes the
field theory model with a detailed presentation found in the appendix.
The third and fourth sections cover the theoretical and empirical
aspects of hedging in a field theory model. The conclusion is left to
section five.

\section{Field Theory} \label{QF}

The field theory model underlying the hedging results in section three
was developed by Baaquie \cite{Baaquie} and calibrated empirically in
Baaquie and Marakani \cite{baaqmar1}. As alluded to earlier, the
fundamental difference between the model presented in this paper and
the original finite factor HJM model stems from the use of a infinite
dimensional random process whose second argument admits correlation between
maturities.

A \textbf{Lagrangian} is introduced to describe the field. The
Lagrangian has the advantage over Brownian motion of being able to
control fluctuations in the field, hence forward rates, with respect
to maturity through the addition of a maturity dependent gradient as
detailed in Definition \ref{L}.  The \textbf{action} of the field
integrates the Lagrangian over time and when exponentiated and
normalized serves as the probability distribution for forward rate
curves. The \textbf{propagator} measures the correlation in the field
and captures the effect the field at time $t$ and maturity $x$ has on
maturity $x^{\prime}$ at time $t^{\prime}$. In the one factor HJM
model, the propagator equals one which allows the quick recovery of one factor HJM results. Previous research by Kennedy \cite{Ken}, Kennedy \cite{Ken1}, and Goldstein \cite{Gold} has begun with the propagator or ``correlation" function for the field instead of deriving this quantity from the Lagrangian. More importantly, the Lagrangian and its associated action generate a path integral that facilitates the solution of contingent claims and hedge parameters. However, previous term structure models have not defined the Lagrangian and are therefore unable to utilize the path integral in their applications. The \textbf{Feynman path integral}, path integral in short, is a fundamental quantity that provides a generating function for forward rate curves and is formally introduced in Definition \ref{pi} of the appendix. Although crucial for pricing and hedging, the path integral has not appeared in previous term structure models with generalized continuous random processes.

\begin{rem}{Notation} \newline
  Let $t_0$ denote the current time and $T$ the set of forward rate
  maturities with $t_0 \leq T$. The upper bound on the forward rate
  maturities is the constant $T_{FR}$ which constrains the forward
  rate maturities $T$ to lie within the interval $\left[ t_0,t_0 +
    T_{FR} \right]$.
\end{rem}

To illustrate the field theory approach, the original finite factor
HJM model is derived using field theory principles in appendix
\ref{details}. In the case of a one factor model, the derivation does
not involve the propagator as the propagator is identically one when
forward rates are perfectly correlated. However, the propagator is non
trivial for field theory models as it governs the imperfect
correlation between forward rate maturities. Let $A(t,x)$ be a two
dimensional field driving the evolution of forward rates $f(t,x)$
through time. Following Baaquie \cite{Baaquie}, the Lagrangian of the field is defined as

\begin{define}{Lagrangian} \newline \label{L}
\noindent The Lagrangian of the field equals

\begin{eqnarray}
\label{LA}
{\cal L}[A] &=& -\frac{1}{2 T_{FR}} \left\{ A^2(t,x)+\frac{1}{\mu^2}
  \left( \frac{\partial A(t,x)}{\partial x} \right)^2 \right\} 
\end{eqnarray}
\end{define}

Definition \ref{L} is not unique, other Lagrangians exist and would
imply different propagators. However, the Lagrangian in Definition
\ref{L} is sufficient to explain the contribution of field theory
while quickly reducing to the one factor HJM framework. Observe the
presence of a gradient with respect to maturity $\displaystyle \frac{\partial A(t,x)}{\partial x}$ that controls field fluctuations in the direction of the forward rate maturity. The constant\footnote{Other functional forms are possible but a constant is chosen for simplicity.} $\mu$ term measures the strength of the fluctuations in the maturity direction. The Lagrangian in Definition \ref{L} implies the field is continuous, Gaussian, and Markovian. Forward rates involving the field are expressed below where the drift and volatility functions satisfy the usual regularity conditions.

\begin{equation}
\label{ndfA} \frac{{\partial f(t,x)}}{{\partial t}} = \alpha(t,x) +\sigma(t,x) A(t,x)
\end{equation}

The forward rate process in equation (\ref{ndfA}) incorporates
existing term structure research on Brownian sheets, stochastic
strings, etc that have been used in previous continuous term structure
models. Note that equation (\ref{ndfA}) is easily generalized to the
$K$ factor case by introducing $K$ independent and identical fields
$A_i(t,x)$. Forward rates could then be defined as

\begin{equation}
\label{dfAi}
\frac{{\partial f(t,x)}}{{\partial t}}=\alpha(t,x) +\sum_{i=1}^K\sigma_i(t,x)A_i(t,x)
\end{equation}

However, a multifactor HJM model can be reproduced without introducing
multiple fields. In fact, under specific correlation functions, the
field theory model reduces to a multifactor HJM model without any
additional fields to proxy for additional Brownian motions.

\begin{prop}{Lagrangian of Multifactor HJM} \label{Con-HJM} \newline
\noindent The Lagrangian describing the random process of a K-factor HJM model is given by

\begin{eqnarray}
{\cal L}[A] &=& - \frac{1}{2} A(t,x) G^{-1}(t,x,x') A(t,x') \nonumber 
\end{eqnarray}

where 

\begin{equation}
\frac{{\partial f(t,x)}}{{\partial t}} = \alpha(t,x) +A(t,x) \nonumber
\end{equation}

and  $G^{-1}(t,x,x')$ denotes the inverse of the function 

\begin{eqnarray}
G(t,x,x') &=& \sum_{i=1}^K \sigma_i(t,x) \sigma_i(t,x') \nonumber
\end{eqnarray} 
\end{prop}

The above proposition is an interesting academic exercises to illustrate the parallel between field theory and traditional multifactor HJM models. However, multifactor HJM models have the disadvantages described earlier in the introduction associated with a finite dimensional basis. Therefore, this approach is not pursued in later empirical work. In addition, it is possible for forward rates to be perfectly correlated within a segment of the forward rate curve but imperfectly correlated with forward rates in other segments. For example, one could designate short, medium, and long maturities of the forward rate curve. This situation is not identical to the multifactor HJM model but justifies certain market practices that distinguish between short,
medium, and long term durations when hedging. However, more complicated correlation functions would be required; compromising model parsimony and reintroducing the same conceptual problems of finite factor models. Furthermore, there is little economic intuition to justify why the correlation between forward rates should be discontinuous. Therefore, this approach is also not considered in later empirical work.

\subsection{Propagator}

The propagator is an important quantity that accounts for the
correlation between forward rates in a parsimonious manner. The
propagator $D(x,x';t,T_{FR})$ corresponding to the Lagrangian in
Definition \ref{L} is given by the following lemma where
$\theta(\cdot)$ denotes a Heavyside function.

\begin{lem}{Evaluation of Propagator} \label{prop} \newline
\noindent The propagator equals 

\begin{eqnarray}
\label{propa}
D(x,x';t,T_{FR})&=&\frac{\mu T_{FR}}{
\sinh(\mu T_{FR})}\Big [\sinh\mu(T_{FR}-\tau)\sinh(\mu\tau')\theta(\tau-\tau')\nonumber\\
&+&\sinh\mu(T_{FR}-\tau')\sinh(\mu\tau)\theta(\tau'-\tau)\nonumber \\
    &+&\frac{1}{2\cosh^2 \left(\frac{\mu T_{FR}}{2} \right)}\Big\{2\cosh\mu\left( \tau- \frac{T_{FR}}{2} \right)
    \cosh\mu \left( \tau'- \frac{T_{FR}}{2} \right)\nonumber\\
    &+&\sinh(\mu \tau) \sinh(\mu\tau')
            +\sinh\mu(T_{FR}-\tau)\sinh\mu(T_{FR}-\tau')\Big\} \Big
            ]\nonumber
\end{eqnarray}

where $\tau=x-t$ and $\tau'=x'-t$ both represent time to maturities.
\end{lem}

Lemma \ref{prop} is proved by evaluating the expectation $E
[A(t,x),A(t',x')]$. The computations are tedious and contained in
Baaquie \cite{Baaquie} but are well known in physics and described in common references such as Zinn-Justin \cite{zj}. The propagator of Goldstein \cite{Gold} is seen as a special case of Lemma \ref{prop} defined on the infinite domain $-\infty < x,x' < \infty$ rather than the finite domain $t \leq x,x' \leq t + T_{FR}$. Hence, the propagator in Lemma \ref{prop} converges to the propagator of Goldstein \cite{Gold} as the time domain expands from a compact set to the real line. The effort in solving for the propagator on the finite domain is justified as it allows covariances near the spot rate $f(t,t)$ to differ from those over longer maturities. Hence, a potentially important boundary condition defined by the spot rate is not ignored. 

Observe that the propagator $D(x,x';t,T_{FR})$ in Lemma \ref{prop} {\it only} depends on the variables $\tau$ and $\tau'$ as well as the correlation parameter $\mu$ which implies that the propagator is time invariant. This important property facilitates empirical estimation in section \ref{Est} when the propagator is calibrated to market data. To understand the significance of the propagator, note that the correlator of the field $A(t,x)$ for $t_0 \leq t,t' \leq t_0+T_{FR}$ is given by

\begin{eqnarray}
\label{aa}
E[A(t,x)A(t',x')]&=&\delta(t-t')D(x,x';t,T_{FR})
\end{eqnarray}

In other words, the propagator measures the effect the value of the
field $A(t,x)$ has on $A(t',x')$; its value at another maturity $x'$ at another point in time. Although $D(x,x';t,T_{FR})$ is complicated in appearance, it collapses to one when $\mu$ equals zero as fluctuations in the $x$ direction are constrained to be perfectly correlated. It is important to emphasize that $\mu$ does not measure the correlation between forward rates. Instead, the propagator solved for in terms of $\mu$ fulfills this role.

\begin{rem}{Propagator, Covariances, and Correlations} \newline
The propagator $D(x,x';t,T_{FR})$ serves as the covariance function for the field while $$\sigma(t,x) D(x,x';t,T_{FR}) \sigma(t,x')$$ serves as the covariance function for forward rates innovations. Hence, the above quantity is repeatedly found in hedging and pricing formulae presented in the next section. The correlation functions for the field and forward rate innovations are identical as the volatility functions $\sigma(t,\cdot)$ are eliminated after normalization. This common correlation function is critical for estimating the field theory model and equals equation (\ref{eq:corr}) of section \ref{Est}. 
\end{rem}

The following table\footnote{The $j(t)$ and $J(t,x)$ functions found in the
definition of the path integral are ``source" functions used to
compute the moments of forward curves (see appendix \ref{details}).
They do not appear in the solution of contingent claims or hedge
parameters.} summarizes the important terms in both the original HJM
model and its extended field theory version.

\vspace{0.3in}

\begin{tabular}{|l|c|c|} \hline
Quantity & Finite Factor HJM & Field Theory  \\ \hline \hline
Lagrangian  & $-\frac{1}{2} W^2(t) $
& $-\frac{1}{2 T_{FR}} \left\{ A^2(t,x)+\frac{1}{\mu^2} \left(
    \frac{\partial A(t,x)}{\partial x} \right)^2 \right\}$ \\ \hline 
Propagator & $1$ & $D(x,x';t,T_{FR})$ \\ \hline
Path Integral &$\exp \left\{ \frac{1}{2}\int_{t_0}^{\infty}dt j^2(t)
\right\}$  & $\exp \left\{
  \frac{1}{2}\int_{t_0}^{\infty}dt\int_t^{t+T_{FR}}dx dx'
  J(t,x)D(x,x';t,T_{FR})J(t,x') \right\}$ \\ \hline 
\end{tabular}

\vspace{0.3in}

As expected, the HJM drift restriction is generalized in the
context of a field theory term structure model. However, producing the
drift restriction follows from the original HJM methodology as the discounted bond price evolves as a martingale under the risk neutral measure to ensure no arbitrage. Under the risk neutral measure, the bond price is written as 

\begin{eqnarray}
P(t_0,T) &=& E_{[t_0,t_*]} \left[ e^{-\int_{t_0}^{t_*}r(t)dt} P(t_*,T) \right] \nonumber \\ &=& \int {\cal D}A e^{-\int_{t_0}^{t_*} dt f(t, t)}
e^{-\int_{t_*}^T dx f(t_*, x)} \label{eq:fb}
\end{eqnarray}

where $\int {\cal D}A$ represents an integral over all possible field paths in the domain $\int_{t_0}^{t_*} dt \int_{t_*}^T dx$. The notation $E_{[t_0,t_*]}[S]$ denotes the expected value under the risk neutral measure of the stochastic variable $S$ over the time interval $[t_0,t_*]$. Equation (\ref{eq:fb}) serves as the foundation for computing the forward rate drift restriction stated in the next proposition and proved in appendix \ref{details}.

\begin{prop}{Drift Restriction} \label{drift} \newline
\noindent The field theory generalization of the HJM drift restriction equals

\begin{eqnarray}
 \alpha(t,x)&=&\sigma(t,x)\int_{t}^{x}dx' D(x,x';t,T_{FR})\sigma(t,x') \nonumber
\end{eqnarray}
\end{prop}

As expected, with $\mu$ equal to zero the result of Proposition \ref{drift} reduces to $$ \alpha(t,x) = \sigma(t,x)\int_{t}^{x}dx'\sigma(t,x')$$ and the one factor HJM drift restriction is recovered. The next section considers the problem of hedging in the context of a field theory model using either bonds or futures contracts on bonds. 

\section{Pricing and Hedging in Field Theory Models} \label{h}

Hedging a zero coupon bond denoted $P(t,T)$ using other zero coupon bonds is accomplished by minimizing the residual variance of the hedged portfolio.  The hedged portfolio $\Pi(t)$ is represented as 

$$\Pi(t) = P(t,T) + \sum_{i=1}^N \Delta_i P(t,T_i)$$ 

where $\Delta_i$ denotes the amount of the $i^{th}$ bond $P(t,T_i)$ included in the hedged portfolio. Note the bonds $P(t,T)$ and $P(t,T_i)$ are determined by observing their market values at time $t$. It is the instantaneous {\it change} in the portfolio value that is stochastic. Therefore, the volatility of this change is computed to ascertain the efficacy of the hedge portfolio. 

For starters, consider the variance of an individual bond in the field theory model. The definition $P(t, T) = \exp{(-\int_t^T dx f(t, x))}$ for zero coupon bond prices implies that 
 
\begin{eqnarray}
\frac{dP(t,T)}{P(t,T)} &=& f(t, t)dt - \int_t^T dx df(t, x) \nonumber \\ &=& \left( r(t)  - \int_t^T dx \alpha(t, x) - \int_t^T dx \sigma(t, x) A(t, x) \right) dt \nonumber
\end{eqnarray}

and $E\left[ \frac{dP(t, T)}{P(t,t)} \right] = \left( r(t) - \int_t^T dx \alpha(t, x) \right) dt$ since $E[A(t, x)] = 0$. Therefore
 
\begin{equation}
\frac{dP(t, T)}{P(t,T)} - E\left[ \frac{dP(t, T)}{P(t,T)} \right] =  -dt\int_t^T dx \sigma(t,x) A(t, x) 
\end{equation}

Squaring this expression and invoking the result that
$E[A(t, x) A(t, x')] = \delta(0) D(x,x';t,T_{FR}) = \frac{D(x,x';t,T_{FR})}{dt}$ results in the instantaneous bond price variance

\begin{equation}
Var[dP(t,T)] = dt P^2(t, T) \int_t^T dx \int_t^T dx' \sigma(t, x) D(x,x';t,T_{FR}) \sigma(t, x')
\end{equation}  

As an intermediate step, the instantaneous variance of a bond portfolio is considered. For a portfolio of bonds, $\hat{\Pi}(t) = \sum_{i=1}^N \Delta_i P(t,T_i)$, the following results follow directly
 
\begin{equation}
d\hat{\Pi}(t) - E[d\hat{\Pi}(t)] = -dt \sum_{i=1}^N \Delta_i P(t,
T_i) \int_t^{T_i} dx \sigma(t, x) A(t, x)
\end{equation}  

and 

\begin{equation}
Var[d\hat{\Pi}(t)] = dt \sum_{i=1}^N
\sum_{j=1}^N \Delta_i \Delta_j P(t,T_i) P(t,T_j) \int_t^{T_i} dx \int_t^{T_j} dx \sigma(t, x) D(x,x';t,T_{FR}) \sigma(t, x') 
\label{eq:bondvariance}
\end{equation}

The (residual) variance of the hedged portfolio $$\Pi(t) = P(t,T) + \sum_{i=1}^N \Delta_i P(t,T_i)$$ may now be computed in a straightforward manner. For notational simplicity, the bonds $P(t,T_i)$ (being used to hedge the original bond) and  $P(t,T)$ are denoted $P_i$ and $P$ respectively. Equation (\ref{eq:bondvariance}) implies the hedged portfolio's variance equals the final result shown below 

\begin{equation}
  \begin{split}
    \label{eq:hedgingbonds}
    &P^2 \int_t^T dx \int_t^T dx' \sigma(t, x) \sigma
    (t, x') D(x,x';t,T_{FR}) \\
    +& 2 P \sum_{i=1}^N \Delta_i P_i \int_t^T dx \int_t^{T_i} dx'
    \sigma(t, x) \sigma (t, x') D(x,x';t,T_{FR})  \\
    +& \sum_{i=1}^N \sum_{j=1}^N \Delta_i \Delta_j P_i P_j
    \int_t^{T_i} dx     \int_t^{T_j} dx' \sigma(t, x) \sigma(t, x')
    D(x,x';t,T_{FR})
  \end{split}
\end{equation}

Observe that the residual variance depends on the correlation between forward rates described by the propagator. Ultimately, the effectiveness of the hedge portfolio is an empirical question since perfect hedging is not possible without shorting the original bond. This empirical question is addressed in section \ref{Est} when the propagator is calibrated to market data. Minimizing the residual variance in equation (\ref{eq:hedgingbonds}) with respect to the hedge parameters $\Delta_i$ is an application of standard calculus. The following notation is introduced for simplicity.

\begin{define} \label{not1}
\begin{eqnarray}
L_i &=& P P_i \int_t^T dx \int_t^{T_i} dx'
\sigma(t, x) \sigma(t, x') D(x,x';t,T_{FR}) \nonumber \\ M_{ij}  &=& P_i
P_j \int_t^{T_i} dx \int_t^{T_j} dx' \sigma(t, x) \sigma(t, x') D(x,x';t,T_{FR})\ \nonumber
\end{eqnarray}
\end{define} 

Definition \ref{not1} allows the residual variance in equation (\ref{eq:hedgingbonds}) to be succinctly expressed as

\begin{eqnarray}
 \label{eq:res}
&&P^2 \int_t^T dx \int_t^T dx' \sigma(t, x) \sigma(t, x') D(x,x';t,T_{FR}) 
+ 2  \sum_{i=1}^N \Delta_i L_i + \sum_{i=1}^N \sum_{j=1}^N \Delta_i \Delta_j M_{ij} 
\end{eqnarray}

Hedge parameters $\Delta_i$ that minimize the residual variance in equation (\ref{eq:res}) are the focus of the next theorem.

\begin{thm}{Hedge Parameter for Bond} \label{dg} \newline
\noindent Hedge parameters in the field theory model equal

\begin{eqnarray}
\Delta_i &=& - \sum_{j=1}^N L_j M_{ij}^{-1} \nonumber 
\end{eqnarray}

and represent the optimal amounts of $P(t,T_i)$ to include in the hedge portfolio when hedging $P(t,T)$.
\end{thm}

Theorem \ref{dg} is proved by differentiating equation (\ref{eq:res}) with respect to $\Delta_i$ and subsequently solving for $\Delta_i$. Corollary \ref{rv} below is proved by substituting the result of Theorem
\ref{dg} into equation (\ref{eq:res}).

\begin{cor}{Residual Variance} \label{rv} \newline
\noindent The variance of the hedged portfolio equals

\begin{eqnarray}
  V &=& P^2 \int_t^T dx \int_t^T dx' \sigma(t, x) \sigma(t,
  x') D(x,x';t,T_{FR}) - \sum_{i=1}^N \sum_{j=1}^N L_i M_{ij}^{-1} L_j \nonumber
\end{eqnarray}

which declines monotonically as $N$ increases.
\end{cor}

The residual variance in Corollary \ref{rv} enables the effectiveness of the hedge portfolio to be evaluated. Therefore, Corollary \ref{rv} is the basis for studying the impact of including different bonds in the hedge portfolio as illustrated in subsection \ref{error}. For $N=1$, the hedge parameter
in Theorem \ref{dg} reduces to

\begin{eqnarray}
\Delta_1 &=& - \frac{P}{P_1} \left( \frac{ \int_t^T dx \int_t^{T_1} dx'
\sigma(t, x) \sigma(t, x') D(x,x';t,T_{FR})}{
 \int_t^{T_1} dx \int_t^{T_1} dx' \sigma(t, x) \sigma(t, x') D(x,x';t,T_{FR})} \right) \label{eq:hed} 
\end{eqnarray}

To obtain the HJM limit, constrain the propagator to equal one. The hedge parameter in equation (\ref{eq:hed}) then reduces to

\begin{eqnarray}
\Delta_1  &=& - \frac{P}{P_1} \left( \frac{ \int_t^T dx \int_t^{T_1} dx'
\sigma(t, x) \sigma(t, x') }{ \left( \int_t^{T_1} dx  \sigma(t, x) \right)^2 } \right) = - \frac{P}{P_1} \left( \frac{ \int_t^T dx \sigma(t, x) }{ \int_t^{T_1} dx  \sigma(t, x)  } \right)\label{eq:nhed} 
\end{eqnarray}

The popular exponential volatility function $\sigma(t,T) = \sigma
e^{-\lambda (T-t)}$ allows a comparison between our field theory
solutions and previous research. Under the assumption of exponential volatility, equation (\ref{eq:nhed}) becomes

\begin{eqnarray}
\Delta_1 &=& - \frac{P}{P_1} \left( \frac{  1-e^{-\lambda(T-t)} }{  1-e^{-\lambda(T_1-t)}}  \right) \label{eq:nhedjt} 
\end{eqnarray}

Equation (\ref{eq:nhedjt}) coincides with the ratio of hedge parameters found as equation 16.13 of Jarrow and Turnbull \cite{JTtext}. In terms of their notation

\begin{eqnarray}
\Delta_1 &=& - \frac{P(t,T)}{P(t,T_1)} \left( \frac{X(t,T)}{X(t,T_1)} \right) \label{eq:jth}
\end{eqnarray}

For emphasis, the following equation holds in a one factor HJM model $$\frac{\partial \left[ P(t,T) + \Delta_1 P(t,T_1) \right]}{\partial r(t)} =0$$ which is verified using equation (\ref{eq:jth}) and results found on pages 494-495 of Jarrow and Turnbull \cite{JTtext} 

\begin{eqnarray}
\frac{\partial \left[ P(t,T) + \Delta_1 P(t,T_1) \right]}{\partial r(t)} &=& 
-P(t,T) X(t,T) - \Delta_1 P(t,T_1) X(t,T_1) \nonumber \\
&=& -P(t,T) X(t,T) + P(t,T) X(t,T) = 0 \nonumber
\end{eqnarray}

When $T_1 =T$, the hedge parameter equals minus one. Economically, this fact states that the best strategy to hedge a bond is to short a bond of the same maturity. This trivial approach reduces the residual variance in equation (\ref{eq:res}) to zero as $\Delta_1=-1$ and $P=P_1$ implies $L_1=M_{11}$. Empirical results for nontrivial hedging strategies are found in subsection \ref{error} after the propagator is calibrated.

\subsection{Futures Pricing}

As this paper is primarily concerned with hedging a bond portfolio, 
futures prices are derived as they are commonly used for hedging bonds given their liquidity.

\begin{prop}{Futures Price} \label{futures} \newline
\noindent The futures price ${\cal F}(t_0, t_*,T)$ is given by 

\begin{eqnarray}
{\cal F}(t_0,t_*,T) &=& E_{[t_0,t_*]}[P(t_*,T)] \nonumber \\
&=& \int {\cal D}A e^{-\int_{t_*}^T dx f(t_*, x)} \nonumber \\
&=& F(t_0,t_*,T) \exp \left\{ \Omega_{\cal F}(t_0,t_*,T) \right\}
\end{eqnarray}

where $ F(t_0,t_*,T)$ represents the forward price for the same contract, 
$F(t_0,t_*,T)=\frac{P(t_0,T)}{P(t_0,t_*)}$, and 

\begin{equation}
\Omega_{\cal F}(t_0,t_*,T)= -\int_{t_0}^{t_*}dt \int_t^{t_*}dx
\sigma(t,x)\int_{t_*}^T dx' D(x,x';t,T_{FR}) \sigma(t,x') \label{eq:om}
\end{equation}
\end{prop}

Details of the proof are found in appendix \ref{details}. Observe that for $\mu =0$, equation (\ref{eq:om}) collapses to 

\begin{equation}
\Omega_{\cal F}(t_0,t_*,T)= -\int_{t_0}^{t_*}dt \int_t^{t_*}dx
\sigma(t,x)\int_{t_*}^T dx' \sigma(t,x') \label{eq:cf}
\end{equation}

which is equivalent to the one factor HJM model. For the one factor HJM model with exponential volatility, equation (\ref{eq:cf}) becomes

\begin{eqnarray}
\Omega_{\cal F}(t_0,t_*,T) &=& -\frac{\sigma^2 }{{2
\lambda^3}} \left(1-e^{-\lambda(T-t_*)} \right)\left(1-e^{-\lambda(t_*- t_0)} \right)^2 \nonumber
\end{eqnarray}

which coincides with equation 16.23 of Jarrow and Turnbull
\cite{JTtext}.  Observe that the propagator modifies the product of
the volatility functions with $\mu$ serving as an additional model parameter. Prices for call options, put options, caps, and floors proceed along similar lines with an identical modification of the volatility functions but their solutions are omitted for brevity. Formulae for these contingent claims are given in Baaquie \cite{Baaquie} where their reduction to the closed form solutions of Jarrow and Turnbull \cite{JTtext} is also presented.

\subsection{Hedging Bonds with Futures Contracts} 

The material in the previous subsection allows the hedging properties of futures contracts on bonds to be studied. Proceeding as before, the appropriate hedge parameters for futures contracts expiring in one year are computed. Proposition \ref{futures} expresses the futures price ${\cal F}(t,t_*,T)$ in terms of the forward price $\frac{P(t, T)}{P(t,
  t_*)} = e^{- \int_{t_*}^T dx f(t,x)}$ and the \textit{deterministic} quantity $\Omega_{\cal F}(t,t_*,T)$ found in equation (\ref{eq:om}). The dynamics of the futures price $d{\cal F}(t,t_*,T)$ is given by

\begin{equation}
  \frac{d{\cal F}(t,t_*,T)}{{\cal F}(t,t_*,T)} = d\Omega_{\cal F}(t,t_*,T) - \int_{t_*}^T dx df(t, x)
\end{equation}

which implies

\begin{equation}
 \frac{ d{\cal F}(t,t_*,T) - E[d{\cal F}(t,t_*,T)]}{{\cal F}(t,t_*,T)} = -dt \int_{t_*}^T dx \sigma(t,x) A(t, x)
\end{equation}

Squaring both sides leads to the instantaneous variance of the futures price  

\begin{equation}
  Var[d{\cal F}(t,t_*,T)] = dt {\cal F}^2(t,t_*,T) \int_{t_*}^T dx \int_{t_*}^T dx' \sigma (t,x) D(x, x) \sigma(t, x')
\end{equation}

The following definition updates Definition \ref{not1} in the context of futures contracts.

\begin{define}{Futures Contracts} \label{not2} \newline
\noindent Let ${\cal F}_i$ denote the futures price ${\cal F}(t,t_*,T_i)$ of a contract expiring at time $t_*$ on a zero coupon bond maturing at time $T_i$. The hedged portfolio in terms of the futures contract is given by
$$\Pi(t) = P + \sum_{i=1}^N \Delta_i {\cal F}_i$$ where ${\cal F}_i$ represent observed market prices. For notational simplicity, define the following terms

\begin{eqnarray}
  \label{eq:M_futures}
        L_i &=& P {\cal F}_i \int_{t_*}^{T_i} dx \int_t^T dx' \sigma (t,
  x) D(x,x';t,T_{FR}) \sigma (t, x') \nonumber \\
  M_{ij} &=& {\cal F}_i{\cal F}_j \int_{t_*}^{T_i} dx \int_{t_*}^{T_j}
  dx' \sigma (t, x) D(x,x';t_*,T_{FR}) \sigma (t, x') \nonumber 
\end{eqnarray}
\end{define}

The hedge parameters and the residual variance when futures contracts are used as the underlying hedging instruments have identical expressions to those in Theorem \ref{dg} and Corollary \ref{rv} but are based on Definition \ref{not2}. Computations parallel those in the beginning of this section.

\begin{cor}{Hedge Parameters and Residual Variance using Futures} \newline
\noindent Hedge parameters for a futures contract that
expires at time $t_*$ on a zero coupon bond that matures at time $T_i$
equals 

\begin{eqnarray}
\Delta_i &=& - \sum_{j=1}^N L_j M_{ij}^{-1} \nonumber 
\end{eqnarray}

while the variance of the hedged portfolio equals

\begin{eqnarray}
  V &=& P^2 \int_t^T dx \int_t^T dx' \sigma(t, x) \sigma(t,
  x') D(x,x';t,T_{FR}) - \sum_{i=1}^N \sum_{j=1}^N L_i M_{ij}^{-1} L_j \nonumber
\end{eqnarray}

for $L_i$ and $M_{ij}$ in Definition \ref{not2}.
\end{cor}

Proof follows directly from previous work.

\section{Empirical Estimation of Field Theory Models} \label{Est}

This section illustrates the significance of correlation between
forward rate maturities and numerically estimates its impact on
hedging performance. The volatility function $\sigma(t,x)$ and the $\mu$ parameter were previously calibrated non parametrically from market data in Baaquie and Srikant \cite{baaqmar1}. The data used in the following empirical tests was generously provided by Jean-Philippe Bouchaud
of Science and Finance. The data consists of daily closing prices for
quarterly Eurodollar futures contracts with a maximum maturity of 7.25 years as described in Bouchaud, Sagna, Cont, and El-Karoui \cite{data} as well as Bouchaud and Matacz \cite{data1}.

As previous empirical research into estimating HJM models has found,
Amin and Morton \cite{Morton}, multifactor HJM models are difficult to
estimate.  This concept is revealed by considering the difference
between a one and two factor HJM model. Let $\tau = x-t$ and $\tau' = x'-t $ once again represent time to maturity and assume the correlation function is time invariant by depending solely on $\tau$ and $\tau'$. The correlation between forward rate maturities $C(x,x')$ may be expressed in terms of the normalized propagator \footnote{The value of $T_{FR}$ was set to $\infty$, rescaling by $\frac{1}{T_{FR}}$ preserved the Lagrangian in Definition \ref{L} and its associated propagator.} as

\begin{eqnarray}
C(x,x') &=& \frac{D(x,x';t)}{\sqrt{D(x,x;t)D(x',x';t)}} 
\label{eq:corr}
\end{eqnarray}

To clarify, the propagator (as emphasized after Lemma \ref{prop}) and hence the correlation in equation (\ref{eq:corr}) is time homogeneous.  
Once the propagator is determined, the correlation between forward rates follows immediately and vice versa. For example, when the propagator equals one, the correlation between forward rates is also identically one. However, for the two factor HJM model, the normalized propagator is given by

\begin{eqnarray}
  \label{eq:twohjm}
  && \frac{1+g(\tau)g(\tau')}{\sqrt{(1+g^2(\tau))(1+g^2(\tau'))}} 
\end{eqnarray}

for $g(\tau)=\frac{\sigma_2(\tau)}{\sigma_1(\tau)}$. Consequently,  
the correlation function above in equation (\ref{eq:twohjm})
depends on the ratio of volatility functions. Therefore, for a given correlation structure, obtaining reliable volatility function estimates is a challenge as it is difficult to disentangle one volatility function from another. This practical disadvantage is overcome by field theory models.

Concerning the estimation of the field theory model, the market volatility function is used throughout the remainder of section \ref{Est}. This volatility function was estimated directly from market data as the variance of term structure innovations for each maturity and is graphed below. 

\begin{figure}[h]
  \centering
  \epsfig{file=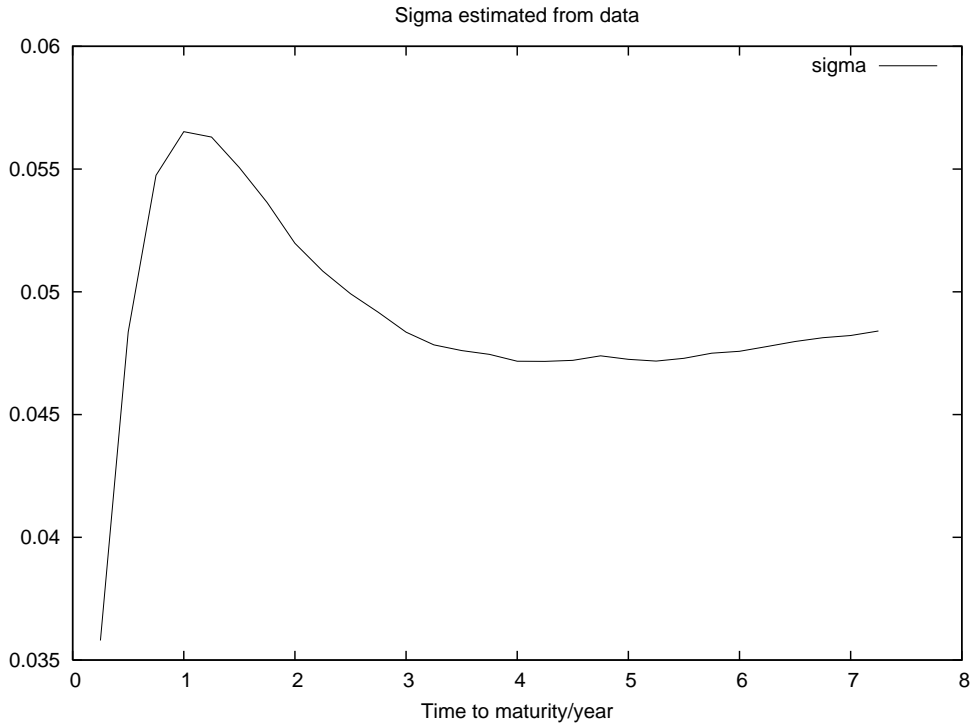, height=14cm, angle=-90}
  \caption{Implied Volatility Function using Market Data}
  \label{fig:marketvol}
\end{figure}

The corresponding implied correlation parameter $\mu$ was estimated as $.06$ (annualized). Estimation of $\mu$ was accomplished after estimating the correlation between forward rate innovations over the 7.25 year horizon. The $\mu$ parameter was then found by minimizing the root mean square difference between the theoretical correlation function in equation (\ref{eq:corr}) which contains $\mu$ via the definition of the propagator in Lemma \ref{prop} and the empirical correlation function generated from the data aggregated over the sample period. The minimization was carried out using the Levenberg-Marquardt method. Stability analysis performed by considering different subsections of the data indicated that this estimate was robust with an error of at most 0.01. The propagator itself is graphed below. 

\begin{figure}[h]
  \centering
  \epsfig{file=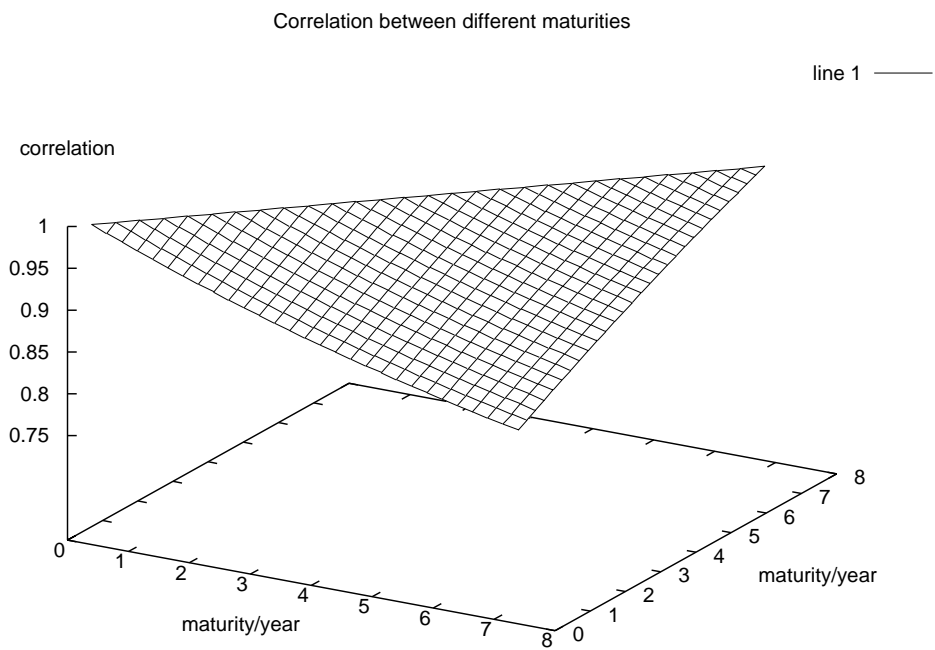, height=14cm, angle=-90}
  \caption{Propagator Implied by $\mu$}
  \label{fig:prop}
\end{figure}

It should be emphasized that the propagator may also be estimated non parametrically from the correlation found in market data without any specified functional form, as the volatility function was estimated. This approach preserves the closed form solutions for hedge parameters and futures contracts illustrated in the previous section. However, the original finite factor HJM model cannot accommodate an empirically determined propagator since it is automatically fixed once the HJM volatility functions are specified.

\subsection{Hedging Error} \label{error}

\begin{figure}[h]
  \centering
  \epsfig{file=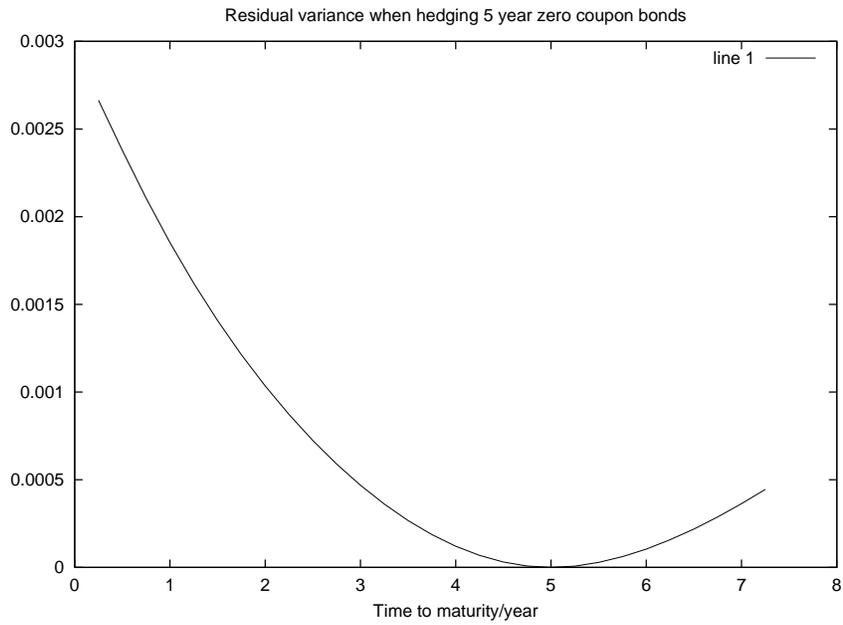, height=12cm, angle=-90}
  \caption{Residual variance for five year bond versus bond maturity used to hedge.}
  \label{fig:resvariance1}
\end{figure}

\begin{figure}[h]
  \centering
  \epsfig{file=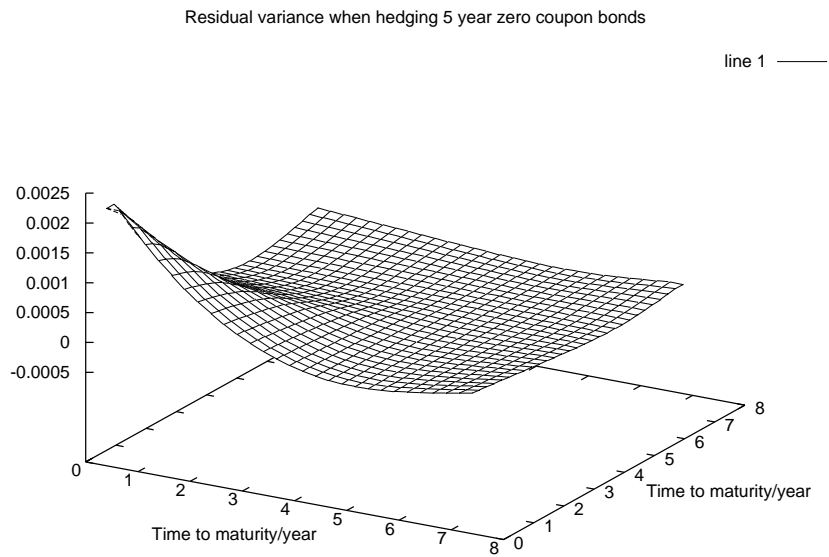, height=12cm, angle=-90}
  \caption{Residual variance for five year bond versus two bond maturities used to hedge.}
  \label{fig:resvariance2}
\end{figure}

\begin{figure}[h]
  \centering
  \epsfig{file=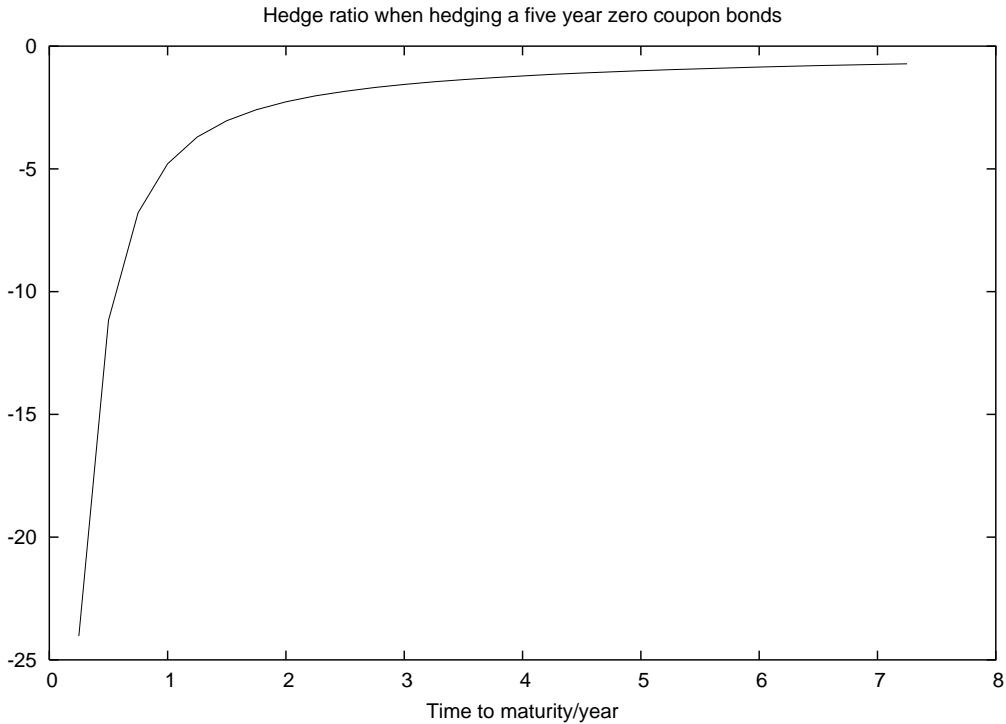, height=14cm, angle=-90}
  \caption{Hedge ratios for five year bond.}
  \label{fig:hedgeratios5yr}
\end{figure}

The reduction in variance achievable by hedging a
five year zero coupon bond with other zero coupon bonds and futures contracts is the focus of this subsection.
The residual variances for one and two bond hedge portfolios are shown in
figures \ref{fig:resvariance1} and \ref{fig:resvariance2}. The
parabolic nature of the residual variance is because $\mu$ is constant. A more complicated function would produce residual variances that do not deviate monotonically as the maturity of the underlying and the hedge portfolio increases although the graphs appeal to our economic intuition which suggests that correlation between forward rates decreases monotonically as the distance between them increases as shown in figure \ref{fig:prop}. Observe that the residual variance drops to zero when the same bond is used to hedge itself; eliminating the original position in the process. The corresponding hedge ratios are shown in figure \ref{fig:hedgeratios5yr}.

\newpage

\newpage

\begin{figure}[h]
  \centering
  \epsfig{file=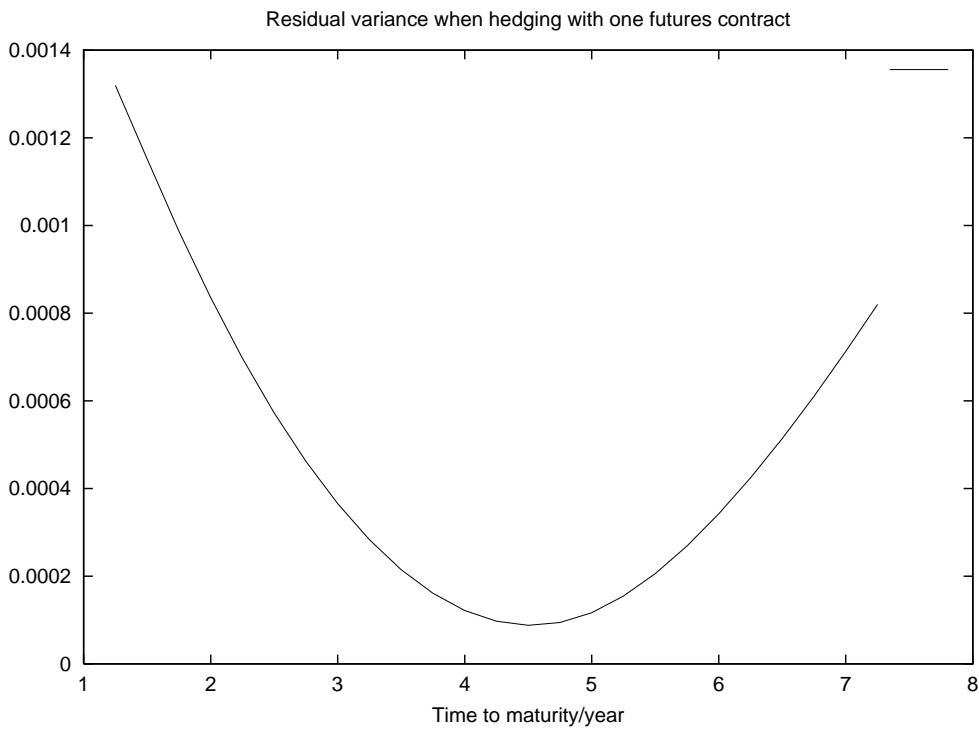, height=14cm, angle=-90}
  \caption{Residual variance for five year bond hedged
    with a one year futures contract on a $T$ maturity bond.}
  \label{fig:resvar_futures}
\end{figure}

Numerical tests to determine the efficiency of hedging using futures
contracts that expire in one year are conducted by calculating the residual variance when hedging a five year zero coupon bond with a futures contract expiring in one year on various zero coupon bonds. The residual variance is
shown in figure \ref{fig:resvar_futures}. Observe that the zero coupon
bond is best hedged by selling futures contracts on 4.5 year bonds
which is explained by the fact that the futures contract only depends on the variation in forward rate curve from $t_*$ to $T$ but the zero coupon bond depends on the variation of the forward rate curve from $t$ to $T$. Hence, a shorter underlying bond maturity is chosen for the futures contract to compensate for the forward rate curve from $t$ to $t_*$. A similar
result is seen when hedging with two futures contracts both expiring in one year. In this case, optimal hedging is obtained when futures contracts on the same bond as well as one on a bond with the minimum possible maturity (1.25 years) are shorted. The use of a futures contract on a short maturity bond is consistent with the high volatility of short maturity forward rates as displayed in figure \ref{fig:marketvol}. The optimal futures contracts to include in the hedge portfolio are shown in table \ref{tab:resvar_futures} when hedging a 5 year bond.

\newpage

\begin{table}[h]
  \centering
  \begin{tabular}[h]{|c|c|c|} \hline
    Number & Futures Contracts (Hedge Ratio) & Residual
    Variance\\ \hline \hline
    0 & none & 0.03046\\ \hline
    1 & $4.5$ years ($-1.288$) & $8.81\times 10^{-5}$\\ \hline
    2 & 5 years ($-0.9347$), $1.25$ years ($-2.72497$) & $2.63 \times
    10^{-5}$\\ \hline
  \end{tabular}
  \caption{Residual variance and hedge ratios for a five year
    bond hedged with one year futures contracts.}
  \label{tab:resvar_futures}
\end{table}

The variances recorded in Table \ref{tab:resvar_futures} correspond to daily dollar denominated fluctuations.

\subsection{Current and Future Research}

The field theory model  has been used to value
call and put options as well as caps and floors in Baaquie
\cite{Baaquie}. Hedge parameters for these instruments is currently under investigation. Obtaining fixed income options data to investigate the significance of correlated forward rates on option prices as well as its impact on their hedge parameters would also be valuable. An enhanced model with stochastic volatility where innovations in the term structure result from the product of two fields has been developed by Baaquie \cite{Baaquiesv}.

\section{Conclusion}

Field theory models address the theoretical limitations of finite factor
term structure models by allowing imperfect correlation between every
forward rate maturity. The field theory model offers computationally expedient hedge parameters for fixed income derivatives and provides a methodology to answer crucial questions concerning the number and maturity of bonds to include in a hedge portfolio. Furthermore, field theory models are able to incorporate correlation between forward rate
maturities in a parsimonious manner that is well suited to empirical
implementation. Empirical evidence revealed the significant impact
that correlation between forward rate maturities has on hedging performance. Despite the infinite dimensional nature of the field, it is
shown that a low dimensional hedge portfolio effectively hedges
interest rate risk by exploiting the correlation between forward
rates. Therefore, field theory models address the theoretical
dilemmas of finite factor term structure models and offer a practical
alternative to finite factor models.

\appendix

\section{Details of Field Theory Model} \label{details}

This appendix briefly reviews the results contained in Baaquie \cite{Baaquie}. The formalism of quantum mechanics is based on conventional
mathematics of partial differential equations and functional analysis
as detailed in Zinn-Justin \cite{zj}. The mathematical tools
underlying quantum field theory have no counterpart in traditional
stochastic calculus although Gaussian fields are equivalent to an
infinite collection of stochastic processes.  The Lagrangian
has the advantage over Brownian motion of being able to control
fluctuations in the field, hence forward rates, with respect to
maturity through the addition of a maturity dependent gradient as
detailed in equation (\ref{nLA}). The action integrates the
Lagrangian over time and when exponentiated and appropriately
normalized yields a probability distribution function for the forward
rate curves. The propagator measures the correlation in the
field and captures the effect the field at time $t$ and maturity $x$
has on the maturity $x^{\prime}$ at time $t^{\prime}$. The
Feynman path integral serves as the generating function for
forward rate curves. The path integral is obtained by integrating the
exponential of the action over all possible evolutions of the
forward rate curve.

\subsection{Restatement of HJM}

Before presenting the field theory of the forward rates, we briefly
restate the HJM model in terms of its original formulation but using
the notation of path integrals. For simplicity, consider a one factor
HJM model of the forward rate curve whose evolution is generated by

\begin{equation}
\label{df}
\frac{{\partial f(t,x)}}{{\partial t}}=\alpha(t,x) + \sigma(t,x) W(t)
\end{equation}

where $\alpha(t,x)$ and $\sigma(t,x)$ are the drift and volatility of
forward rates. For every value of time $t$, the stochastic variable
$W(t)$ is an independent Gaussian random variable with the property that

\begin{equation}
E[W(t)W(t')]= \delta(t-t') \label{eq:verify}
\end{equation}

in contrast to equation (\ref{aa}) involving the field. More conventionally, the HJM model is written as $df (t, x) = \alpha
(t, x) dt + \sigma(t, x) dZ(t)$ where $Z(t)$ represents a Brownian motion
process. Hence, the Gaussian process $W(t)$ equals
$\frac{dZ(t)}{dt}$. To derive the covariance function implied for $W(t)$, the correlation function for $Z(t)$ must be twice differentiated. Therefore, $\langle W(t) W(t')\rangle = \frac{\partial^2}{\partial t \partial t'} \langle Z(t) Z(t') \rangle = \frac{\partial^2 \min(t, t')}{\partial t \partial t'} = \frac{\partial \, \theta (t-t')}{\partial t} = \delta(t-t')$ as seen in equation (\ref{eq:verify}). The Lagrangian of the Gaussian process is defined as

\begin{eqnarray}
\label{LW} {\cal L}[W] &=& - \frac{1}{2} W^2(t) 
\end{eqnarray}

The Gaussian process may be illustrated by discretizing time into a discrete lattice of spacing $\epsilon$, and $\displaystyle {t \rightarrow m\epsilon}$, with $m=1,2....M$, with $W(t) \rightarrow W(m)$. The probability measure underlying Gaussian paths for $t_1 < t < t_2$ is given by

\begin{eqnarray}
\label{PD}
{\cal P}[W]&=& \prod_{m=1}^{M}e^{-\frac{\epsilon}{2} W^2(m)}
= \prod_{m=1}^{M}e^{\epsilon {\cal L}[W]} = e^{\epsilon \sum_{m=1}^M{\cal L}[W]} \\
\int dW&=&\prod_{m=1}^{M}\sqrt{\frac{\epsilon}{2\pi}}
\int_{-\infty}^{+\infty} dW(m) \nonumber
\end{eqnarray}

The term ${\cal P}[W]$ represents the probability of a path traced out by $W(t)$. For purposes of rigor, the continuum notation represents
taking the continuum limit of the discrete multiple integrals given
above.  The infinite dimensional integration measure given by
$\displaystyle{\prod_{t_1<t<t_2}\int_{-\infty}^{+\infty} dx(t)}$ has a
rigorous, measure theoretic, definition as the integration over all
continuous, but nowhere differentiable, paths running between points
$x(t_1)$ and $x(t_2)$. For $t_1<t<t_2$, the probability distribution
function for the paths of the Gaussian process equals

\begin{eqnarray}
\label{P}
{\cal P}[W,t_1,t_2]= \exp \left\{ S[W,t_1,t_2] \right\}
\end{eqnarray}

where the action $S[W,t_1,t_2]$ for the Gaussian process is given by the limit of the exponent in equation (\ref{PD}) as $\epsilon \rightarrow 0$

\begin{eqnarray}
\label{s0}
S[W,t_1,t_2]=-\frac{1}{2} \int_{t_1}^{t_2} W^2(t) dt
\end{eqnarray}

The term $\int {\cal D}W$ denotes path integration over all the random
variables $W(t)$ which appear in the problem. A path integral approach
to the HJM model has been discussed in Chiarella and El-Hassan
\cite{cc} although the action derived is different than the one given
above since a different set of variables were involved. A formula for
the generating function of forward rates driven by a Gaussian process is
given by the path integral

\begin{eqnarray}
\label{z}
Z[j,t_1,t_2]&=&\int {\cal D} W e^{\int_{t_1}^{t_2}dt
    j(t)W(t)}e^{S[W,t_1,t_2]}\nonumber  \\
    &=&\exp \left\{ \frac{1}{2}\int_{t_1}^{t_2} j^2(t) dt \right\}
\end{eqnarray}

This path integral is crucial for applications involving the pricing
of derivatives as demonstrated in section \ref{h}. 

\subsection{Field Theory Model}

 The Lagrangian of the field equals

\begin{equation}
\label{nLA}
{\cal L}[A] = -\frac{1}{2 T_{FR}} \left\{ A^2(t,x)+\frac{1}{\mu^2}
  \left( \frac{\partial A(t,x)}{\partial x} \right)^2 \right\} 
\end{equation}

Since the Lagrangian is quadratic, the resulting field is Gaussian and
the existence of a Hamiltonian for the Lagrangian implies the field is
Markovian. Forward rates are expressed as

\begin{equation}
\label{dfA} \frac{{\partial f(t,x)}}{{\partial t}} = \alpha(t,x) +\sigma(t,x) A(t,x)
\end{equation}

The action of $S[A]$ integrates the Lagrangian over time to yield

\begin{eqnarray}
\label{sa}
S[A] &=& \int_{t_0}^{\infty}dt \int_t^{t+T_{FR}}
dx {\cal L}[A] \nonumber \\ &=& -\frac{1}{2 T_{FR}}
\int_{t_0}^{\infty}dt \int_t^{t+T_{FR}} dx \left\{
  A^2(t,x)+\frac{1}{\mu^2} \left( \frac{\partial A(t,x)}{\partial x}
  \right)^2 \right\} 
\end{eqnarray}

Consider the one factor case

\begin{equation}
\label{wa} W(t) = \int_t^{t+T_{FR}} A(t,x) dx
\end{equation}

by assuming the correlation $\mu$ is zero. This process reduces the
action found in equation (\ref{sa}) to the following action

\begin{eqnarray}
S[A] &=&  -\frac{1}{2 T_{FR}} \int_{t_0}^{\infty} W^2(t) dt
\int_t^{t+T_{FR}} dx \nonumber \\ &=& -\frac{1}{2}\int_{t_0}^{\infty}
W^2(t) dt 
\end{eqnarray}

which is identical to equation (\ref{s0}). If one thinks of the field
$A(t_0,x)$ at some instant $t_0$ as the position of a string, then the
one factor HJM model constrains the string to be rigid. The action
$S[A]$ given in (\ref{sa}) allows every maturity $x$ in $A(t_0,x)$ to
fluctuate as a string with string rigidity equal to $\frac{1}{\mu^2}$. For $\mu=0$, the string is infinitely rigid or perfectly correlated.
A normalizing constant $Z$ is the result of integrating the paths of
the field against the probability of each path.

\begin{equation}
\label{zaa} Z = \int {\cal D} A e^{S[A]}
\end{equation}

where the notation $\int {\cal D} A$ represents an integral over all possible field paths contained in the domain. The moment generating functional for the field is given by the Feynman path integral, Zinn-Justin
\cite{zj}.

\begin{define}{Path Integral} \label{pi} \newline
  The path integral over all possible paths of the field, weighted by
  their probability, equals

\begin{eqnarray}
Z[J] &=& \frac{1}{Z}\int {\cal D} A e^{\int JA} e^{S[A]} \nonumber
\end{eqnarray}

where $Z$ is defined in equation (\ref{zaa}). 
\end{define}

In the context of the present term structure model, the path integral equals

\begin{eqnarray}
Z[J] &=& \frac{1}{Z} \int {\cal D}A \exp \left\{ \int_{t_0}^{\infty} dt 
    \int_t^{t+T_{FR}} dx J(t,x)A(t,x) \right\}  e^{S[A]}
\label{eq:Z} 
\end{eqnarray}

The above formula is interpreted as follows, the $Z$ term operates as
a normalizing constant while $\frac{e^{S[A]}}{Z}$ represents the
probability distribution function for each random path generated by
the field $A(t,x)$. The integrand $$\exp \left\{ \int_{t_0}^{\infty}
  dt \int_t^{t+T_{FR}} dx J(t,x)A(t,x) \right\}$$
contains an external source\footnote{To make an analogy with
  univariate generating functions, the function $J(t,x)$ serves a
  similar role to $t$ in the moment generating function $e^{\mu t +
    \frac{1}{2} \sigma^2 t^2}$ for normal random variables
  $\mathcal{N}(\mu,\sigma^2)$.} function $J(t,x)$ coupled to the field
$ A(t,x)$ that is operated on to produce moments of the distribution.
The path integral $Z[J]$ is evaluated explicitly in Baaquie
\cite{Baaquie} and simplifies equation (\ref{eq:Z}) to

\begin{equation}
\label{za}
Z[J]=\exp \left\{ \frac{1}{2}\int_{t_0}^{\infty}dt\int_t^{t+T_{FR}}dx
  dx' J(t,x)D(x,x';t,T_{FR})J(t,x') \right\} 
\end{equation}

The path integral is the basis for all applications of term structure
modeling such as the pricing and hedging of fixed income contingent
claims since it represents the generating function for forward rate
curves.


\subsection{Proof of Proposition \ref{drift}}

Starting from equation (\ref{eq:fb})

\begin{eqnarray}
P(t_0,T) &=& \int {\cal D}A e^{-\int_{t_0}^{t_*} dt f(t, t)}
e^{-\int_{t_*}^T dx f(t_*, x)} \nonumber 
\end{eqnarray}

and using the identity $P(t_0, T) = e^{-\int_{t_0}^T dx f(t_0, x)}$ implies

\begin{eqnarray}
e^{-\int_{t_0}^T dx f(t_0, x)} &=& \int {\cal D}A e^{-\int_{t_0}^{t_*} dt
  \int_{t_0}^{t} dt' \alpha (t', t) + \sigma (t', t) A(t', t)} e^{-\int_{t_0}^{t_*} dt   \int_{t_*}^T dx \alpha(t, x) + \sigma (t, x) A(t, x)}\nonumber \\ 1 &=& \int {\cal D}A e^{-\int_{t_0}^{t_*} dt \int_t^T dx
  \alpha (t, x) + \sigma (t,x) A(t, x)} \nonumber \\
1 &=& e^{-\int_{t_0}^{t_*} dt \int_t^T dx \alpha  (t, x) + \frac{1}{2}
    \int_{t_0}^{t_*} dt \int_t^T dx \int_t^T dx' \sigma(t, x) D(x,x';t,T_{FR})
    \sigma(t, x')} \nonumber
\end{eqnarray}

and isolates the drift in the exponent as

\begin{eqnarray}
\int_{t_0}^{t_*} dt \int_t^T dx \alpha(t,x) &=& \frac{1}{2}
\int_{t_0}^{t_*} dt \int_t^T dx \int_t^T dx' \sigma(t, x) D(x,x';t,T_{FR})
\sigma(t, x')
\end{eqnarray}

and consequently

\begin{equation}
\label{ad}
\int_{t}^{T}dx\alpha(t,x)=\frac{1}{2}\int_{t}^{T}dxdx'\sigma(t,x)D(x,x';t,T_{FR}) \sigma(t,x')
\end{equation}

This no arbitrage condition must hold for any Treasury bond maturing at
any time $x=T$. Hence, differentiating the above expression with
respect to $T$ produces the final version of the drift restriction

\begin{eqnarray}
\label{noara}
\alpha(t,x)&=&\sigma(t,x)\int_{t}^{x}dx' D(x,x';t,T_{FR})\sigma(t,x')
\end{eqnarray}

\subsection{Proof of Proposition \ref{futures}}

\begin{eqnarray}
{\cal F}(t_0,t_*,T) &=& \int {\cal D}A e^{-\int_{t_*}^T dx f(t_*, x)} \nonumber \\ &=& e^{-\int_{t_*}^T dx f(t_0, x)} \int {\cal D}A
e^{-\int_{t_0}^{t_*}dt \int_{t_*}^T dx \alpha(t, x) + \sigma (t, x)
  A(t, x)} \nonumber \\
&=& F(t_0, t_*, T) e^{-\int_{t_0}^{t_*} dt \int_{t_*}^T dx \alpha(t, x)}
  Z[\sigma(t, x)\theta(x-t_*)\theta(T-x)]\nonumber
\end{eqnarray}

where $Z[J]$ is defined in equation (\ref{eq:Z}). Proceeding further leads to

\begin{eqnarray}
{\cal F}(t_0,t_*,T)&=& F(t_0, t_*, T) e^{-\int_{t_0}^{t_*} dt \int_{t_*}^T dx \int_t^x dx'
  \sigma (t, x) D(x,x';t,T_{FR}) \sigma(t, x')} e^{\frac{1}{2}
  \int_{t_0}^{t_*} dt \int_{t_*}^T dx dx' \sigma(t, x) D(x,x';t,T_{FR})
  \sigma(t, x')}  \nonumber \\
&=& F(t_0,t_*,T) \exp \left\{ \Omega_{\cal F}(t_0,t_*,T) \right\}
\end{eqnarray}

where $\Omega_{\cal F}(t_0,t_*,T)$ is given by equation (\ref{eq:om}).

\bibliography{list-int}

\begin{thebibliography}{10}

\bibitem{Baaquie}
Belal~E. Baaquie.
\newblock Quantum {F}ield {T}heory of {T}reasury {B}onds.
\newblock {\em Physical Review}, 64:1--16, 2001.

\bibitem{Baaquiesv}
Belal~E. Baaquie.
\newblock Quantum {F}ield {T}heory of {F}orward {R}ates with {S}tochastic
  {V}olatility.
\newblock Forthcoming in Physical Review
  http://xxx.lanl.gov/abs/cond-mat/0110506, 2002.

\bibitem{baaqmar1}
Belal~E. Baaquie and Marakani Srikant.
\newblock Empirical {I}nvestigation of a {Q}uantum {F}ield of {F}orward
  {R}ates.
\newblock National University of Singapore
  http://xxx.lanl.gov/abs/cond-mat/0106317, 2002.

\bibitem{Con}
Tomas Bjork and Bent~Jesper Christensen.
\newblock Interest {R}ate {D}ynamics and {C}onsistent {F}orward {R}ate
  {C}urves.
\newblock {\em Mathematical Finance}, 9(4):323--348, October 1999.

\bibitem{BKR}
Tomas Bjork, Yuri Kabanov, and Wolfgang Runggaldier.
\newblock Bond {M}arket in the {P}resence of {M}arked {P}oint {P}rocesses.
\newblock {\em Mathematical Finance}, 7(2):211--239, 1997.

\bibitem{data}
J.P. Bouchaud, N.~Sagna, R.~Cont, N.~El-Karoui, and M.~Potters.
\newblock Phenomenology of the {I}nterest {R}ate {C}urve.
\newblock Working Paper http://xxx.lanl.gov/cond-mat/9712164, 1997.

\bibitem{cc}
C.~Chiarella and N.~El-Hassan.
\newblock Evaluation of {D}erivative {S}ecurity {P}rices in the
  {H}eath-{J}arrow-{M}orton {F}ramework as {P}ath {I}ntegrals {U}sing {F}ast
  {F}ourier {T}ransform {T}echniques.
\newblock {\em Journal of Financial Engineering}, 6(2):121--147, 1997.

\bibitem{Cohen}
Jason Cohen and Robert Jarrow.
\newblock Markov {M}odeling in the {H}eath, {J}arrow, and {H}eath {T}erm
  {S}tructure {F}ramework.
\newblock Cornell University, 2000.

\bibitem{Gold}
Robert Goldstein.
\newblock The {T}erm {S}tructure of {I}nterest {R}ate {C}urve as a {R}andom
  {F}ield.
\newblock {\em Review of Financial Studies}, 13(2):365--384, 2000.

\bibitem{HJM}
Robert Jarrow, David Heath, and Andrew Morton.
\newblock Bond {P}ricing and the {T}erm {S}tructure of {I}nterest {R}ates: {A}
  {N}ew {M}ethodology for {C}ontingent {C}laims.
\newblock {\em Econometrica}, 60(1):77--105, January 1992.

\bibitem{JTtext}
Robert Jarrow and Stuart Turnbull.
\newblock {\em Derivative Securities, Second Edition}.
\newblock South Western College Publishing, 2000.

\bibitem{Ken}
D.P. Kennedy.
\newblock The {T}erm {S}tructure of {I}nterest {R}ates as a {G}aussian {F}ield.
\newblock {\em Mathematical Finance}, 4(3):247--258, July 1994.

\bibitem{Ken1}
D.P. Kennedy.
\newblock Characterizing {G}aussian {M}odels of the {T}erm {S}tructure of
  {I}nterest {R}ates.
\newblock {\em Mathematical Finance}, 7(2):107--118, April 1997.

\bibitem{data1}
Andrew Matacz and Jean-Philippe Bouchaud.
\newblock An {E}mpirical {I}nvestigation of the {F}orward {I}nterest {R}ate
  {T}erm {S}tructure.
\newblock {\em International Journal of Theoretical and Applied Finance},
  3(4):703--729, 2000.

\bibitem{Morton}
Andrew Morton and Kaushik Amin.
\newblock Implied {V}olatility {F}unctions in {A}rbitrage {F}ree {T}erm
  {S}tructure {M}odels.
\newblock {\em Journal of Financial Economics}, 35:141--180, 1994.

\bibitem{String}
Pedro Santa-Clara and Didier Sornette.
\newblock The {D}ynamics of the {F}orward {I}nterest {R}ate {C}urve with
  {S}tochastic {S}tring {S}hocks.
\newblock {\em Review of Financial Studies}, 14(1):149--185, 2001.

\bibitem{zj}
J.~Zinn-Justin.
\newblock {\em Quantum Field Theory and Critical Phenomenon}.
\newblock Cambridge University Press, 1992.

\end{thebibliography}
\bibliographystyle{plain}

\end{document}